\documentclass[12pt,thmsa]{article}
\usepackage{amssymb}

\usepackage{sw20lart}

\input{tcilatex}
\begin{document}

\author{Emilio Santos \and Departamento de F\'{i}sica. Universidad de Cantabria.
Santander. Spain}
\title{Dark energy and quantum vacuum fluctuations}
\date{February, 25, 2009 }
\maketitle

\begin{abstract}
It is suggested that the vacuum expectation of the quantum vacuum
energy-momentum is zero, but quantum fluctuations give rise to a space-time
curvature equivalent to that of a cosmological constant or dark energy.
Calculations within quantized gravity, following a few plausible hypotheses,
provide results compatible with cosmological observations.

PACS: 04.60.-m; 98.80.Hw

\textit{Keywords: }Dark energy; Vacuum fluctuations; Quantum gravity.
\end{abstract}

The observed accelerated expansion of the universe\cite{Sahni} is assumed to
be due to a positive mass density and negative pressure, constant throughout
space and time, which is popularly known as ``dark energy''. The mass (or
energy) density, $\rho _{DE},$ and the presure, $p_{DE},$ are \cite{WMAP} 
\begin{equation}
\rho _{DE}\simeq -p_{DE}\simeq 10^{-26}kg/m.  \label{1}
\end{equation}
(Throughout this paper I shall use units $c=
\rlap{\protect\rule[1.1ex]{.325em}{.1ex}}h%
=1$, but write explicitly Newton\'{}s constant, $G$, for the sake of
clarity.)

The current wisdom is to identify the dark energy with the cosmological
constant introduced by Einstein in 1917 or, what is equivalent in practice,
to assume that it corresponds to the quantum vacuum. Indeed the equality $%
\rho _{DE}=-p_{DE}$ is appropriate for the vacuum (in Minkowski space, or
when the space-time curvature is small) because it is invariant under
Lorentz transformations. A problem appears however when one attempts to
estimate the value of $\rho _{DE}.$ In fact if the dark energy is really due
to the quantum vacuum it seems difficult to understand why the mass density
is not either strictly zero or of the order of Planck\'{}s density, that is 
\begin{equation}
\rho _{DE}\sim \frac{c^{5}}{G^{2}
\rlap{\protect\rule[1.1ex]{.325em}{.1ex}}h%
}\simeq 10^{97}kg/m,  \label{2}
\end{equation}
which is about 123 orders larger than the observed eq.$\left( \ref{1}\right) 
$.

In this paper I explore the possibility that the quantum vacuum energy is
indeed zero but the quantum fluctuations give rise to a curvature of
space-time similar to the one produced by a constant classical
(non-fluctuating) mass density and pressure as given by eq.$\left( \ref{1}%
\right) .$ More correctly stated, the hypothesis is that the quantum vacuum
consists a set of interacting relativistic quantum fields giving rise to an
energy-momentum quantum tensor operator, $\widehat{T}_{\mu }^{\nu }\left(
x^{\eta }\right) ,$ whose vacuum expectation is zero at any space-time
point. That is 
\begin{equation}
\left\langle 0\left| \widehat{T}_{\mu }^{\nu }\left(
x^{1},x^{2},x^{3},x^{4}\right) \right| 0\right\rangle =0,  \label{0c}
\end{equation}
where $\mid 0\rangle $ is the state-vector of the vacuum and $%
x^{1},x^{2},x^{3},x^{4}$ the coordinates of a space-time point, which I
shall label collectively $x^{\eta }$ in the following$.$ In contrast the
existence of quantum fluctuations implies that the vacuum expectation of the
product of the components at two space-time points may not be zero, that is 
\begin{equation}
\left\langle 0\left| \widehat{T}_{\mu }^{\nu }\left( x^{\eta }\right) 
\widehat{T}_{\sigma }^{\lambda }\left( x^{\zeta }\right) \right|
0\right\rangle \neq 0.  \label{0d}
\end{equation}

The effect of the quantum vacuum on the curvature of spacetime should be
calculated within the framework of quantized gravity. This means assuming
that the vacuum is characterized by a metric tensor operator $\widehat{g}%
_{\mu \nu }\left( x^{\eta }\right) $ which is related to the energy-momentum
tensor operator $\widehat{T}_{\mu }^{\nu }\left( x^{\eta }\right) $ by some
equations to be specified. An obvious constraint on these equations is that
they will agree with Einstein\'{}s equations in the classical limit.

My aim is to see whether the quantum vacuum fluctuations, plus the matter
content of the universe, may give rise to the same spce-time curvature as in
the standard model. The most common description of space-time in cosmology
involves the use of the Robertson-Walker-Friedman metric. However for our
purposes it is more convenient to use a local frame with ``curvature''
coordinates appropriate for spherical symmetry around some arbitrary point
of space, that is 
\begin{equation}
ds^{2}=A\left( r,t\right) dr^{2}+r^{2}d\theta ^{2}+r^{2}\sin ^{2}\theta
d\phi ^{2}-B\left( r,t\right) dt^{2}.  \label{ds}
\end{equation}
For a relatively small region around the origin of the coordinate system the
expressions of $A$ and $B$ are simple, namely\cite{Rich} 
\begin{eqnarray}
A &=&1+\frac{8\pi }{3}Gr^{2}\left( \rho _{mat}+\rho _{DE}\right) +O\left(
r^{4}\right) ,  \nonumber \\
B &=&1+\frac{8\pi }{3}Gr^{2}\left( \frac{1}{2}\rho _{mat}-\rho _{DE}\right)
+O\left( r^{4}\right) ,  \label{metric}
\end{eqnarray}
where $\rho _{mat}$ is the density of cold matter, either baryonic or dark, $%
\rho _{DE}$ is the density of dark energy as given in eq.$\left( \ref{1}%
\right) $ and I ignore the (small) contributions of radiation and cold
matter pressure. Thus the task is to reproduce eqs.$\left( \ref{metric}%
\right) $ without a real dark energy density, but including the effect of
quantum vacuum fluctuations.

Although a complete quantum gravity theory, not yet available, would be
needed for a rigorous treatment, we may derive some relevant results via
introducing a few plausible hypotheses. For the sake of clarity I will write
explicitly these assumptions as ``propositions''.

\begin{proposition}
The global properties of space-time, e.g. the accelerated expansion of the
universe or the mean curvature of space if any, may be obtained from the
vacuum expectation value of the metric tensor operator, that expectation
being treated as if it was an actual $classical$ metric tensor.
\end{proposition}

That is I will assume that the following (classical, c-number) metric tensor 
\begin{equation}
g_{\mu \nu }\left( x^{\eta }\right) =\left\langle 0\left| \widehat{g}_{\mu
\nu }\left( x^{\eta }\right) \right| 0\right\rangle ,  \label{0f}
\end{equation}
determines the global properties of space-time. Obviously the quantum
fluctuations of the metric cannot be derived from $g_{\mu \nu }$. In
particular 
\[
\left\langle 0\left| \widehat{g}_{\mu \nu }\left( x^{\eta }\right) \widehat{g%
}_{\lambda \sigma }\left( x^{\zeta }\right) \right| 0\right\rangle \neq
g_{\mu \nu }\left( x^{\eta }\right) g_{\lambda \sigma }\left( x^{\zeta
}\right) .
\]
In order to make a comparison with eqs.$\left( \ref{ds}\right) $ and $\left( 
\ref{metric}\right) ,$ I will write eqs.$\left( \ref{0f}\right) $ in
curvature coordinates. Thus they are specified as follows.

\begin{proposition}
The vacuum expectation values of the diagonal metric coefficients are 
\begin{eqnarray}
\left\langle 0\left| \widehat{g}_{11}\left( x^{\eta }\right) \right|
0\right\rangle  &=&A(r,t),\;\left\langle 0\left| \widehat{g}_{22}\left(
x^{\eta }\right) \right| 0\right\rangle =r^{2},  \nonumber \\
\left\langle 0\left| \widehat{g}_{33}\left( x^{\eta }\right) \right|
0\right\rangle  &=&r^{2}\sin ^{2}\theta ,\;\left\langle 0\left| \widehat{g}%
_{44}\left( x^{\eta }\right) \right| 0\right\rangle =B(r,t),  \label{10}
\end{eqnarray}
the vacuum expectations of non-diagonal coefficients being zero.
\end{proposition}

These relations are a consequence of our choice of coordinates plus the
assumption that the distribution of matter is isotropic on the large scale,
a standard approximation in cosmology.

Now I will state the energy-momentum content of the universe as follows.

\begin{proposition}
In addition to the contribution of the quantum vacuum, there is a cold
matter density, $\rho _{mat},$ (either baryonic or dark) which is uniform in
space but depends on time. Thus the total density and pressure operators are 
\begin{equation}
\widehat{\rho }\left( \mathbf{x,}t\right) =\widehat{I}\rho _{mat}\left(
t\right) +\widehat{\rho }_{vac}\left( \mathbf{x,}t\right) ,\;\widehat{p}%
\left( \mathbf{x,}t\right) =\widehat{p}_{vac}\left( \mathbf{x,}t\right) ,
\label{32a}
\end{equation}
where $\widehat{I}$ is the identity operator and $\mathbf{x}$\textbf{\ }is
the vector with polar coordinates $\left( r,\theta ,\phi \right) .$
\end{proposition}

For the sake of simplicity I neglect the small contributions of the hot
matter (radiation) and the pressure associated to cold matter.

The next task will be to relate the coefficients $A$ and $B$ with the
distribution of matter plus the quantum vacuum fluctuations. As a guide I
shall start from relations valid in classical gravity for a space-time of
spherical symmetry in curvature coordinates. These relations are\cite{Synge} 
\begin{eqnarray}
A\left( r,t\right) &=&\left( 1-\frac{2Gm\left( r\right) }{r}\right) ^{-1},%
\text{ }m\left( r\right) \equiv 4\pi \int_{0}^{r}\rho \left( x\right)
x^{2}dx,  \nonumber \\
B(r,t) &=&\exp \left[ 2G\int_{0}^{r}\frac{m\left( x\right) +4\pi
x^{3}p\left( x\right) }{x^{2}-2Gm\left( x\right) x}dx\right] ,  \label{29}
\end{eqnarray}
$\rho $ being the density and $p$ the pressure of a perfect fluid.

For the purposes of this paper it is enough to work to second order in
Newton\'{}s constant, G. Thus I will write the first eq.$\left( \ref{29}%
\right) $ in the form 
\begin{equation}
A\left( r,t\right) =1+\frac{2Gm}{r}+\frac{4G^{2}m^{2}}{r^{2}}+O\left(
G^{3}\right) .  \label{28}
\end{equation}
Similarly the second eq.$\left( \ref{29}\right) $ may be written 
\begin{eqnarray}
B(r,t) &=&1+2G\int_{0}^{r}\left[ x^{-2}m(x)+4\pi xp\left( x\right) \right] dx
\nonumber \\
&&+4G^{2}\int_{0}^{r}\left[ x^{-3}m(x)^{2}+4\pi m(x)p(x)\right] dx  \nonumber
\\
&&+2G^{2}\left[ \int_{0}^{r}\left[ x^{-2}m(x)+4\pi xp\left( x\right) \right]
dx\right] ^{2}+O\left( G^{3}\right) .  \label{31}
\end{eqnarray}

In order to pass to quantized gravity we should write $A$ and $B$ as vacuum
expectations of expressions involving the operators $\widehat{\rho }\left( 
\mathbf{x,}t\right) \;$and $\widehat{p}\left( \mathbf{x,}t\right) $
introduced in eqs.$\left( \ref{32a}\right) .$ In the absence of any clear
hint, I might suppose that those expressions are similar to eqs.$\left( \ref
{29}\right) $ and $\left( \ref{31}\right) .$ Thus I will assume the
following.

\begin{proposition}
The vacuum expectations of the metric coefficients are related to the
quantum operators of density, $\widehat{\rho },$ and pressure, $\widehat{p},$
(the diagonal elements of the energy-momentum tensor operator $\widehat{T}%
_{\mu }^{\nu }\left( x^{\eta }\right) )$ by
\end{proposition}

\begin{eqnarray}
A\left( r,t\right) &\simeq &1+\left\langle 0\left| \frac{2G}{r}\widehat{m}(r)%
\mathbf{+}\frac{4G^{2}}{r^{2}}\widehat{m}(r)^{2}\right| 0\right\rangle ,%
\text{ }  \nonumber \\
\widehat{m}(r) &\equiv &\int_{\left| \mathbf{x}\right| \leq r}\widehat{\rho }%
\left( \mathbf{x}\right) d^{3}\mathbf{x}  \label{31b}
\end{eqnarray}
\begin{eqnarray}
B(r,t) &\simeq &1+\frac{2G}{r}\left\langle 0\left| \int_{0}^{r}x^{-2}%
\widehat{m}(x)dx+\int_{\left| \mathbf{x}\right| \leq r}x^{-1}\widehat{p}%
\left( \mathbf{x}\right) d^{3}\mathbf{x}\right| 0\right\rangle +B_{2}(r,t) 
\nonumber \\
B_{2}(r,t) &=&2G^{2}\left\langle 0\left| 2\int_{0}^{r}x^{-3}\widehat{m}%
(x)^{2}dx+\int_{\left| \mathbf{x}\right| \leq r}x^{-2}\left[ \widehat{m}(x)%
\widehat{p}\left( \mathbf{x}\right) +\widehat{p}\left( \mathbf{x}\right) 
\widehat{m}(x)\right] d^{3}\mathbf{x}\right| 0\right\rangle  \nonumber \\
&&+2G^{2}\left\langle 0\left| \left[ \int_{0}^{r}x^{-2}\widehat{m}(x)dx%
\mathbf{+}\int_{\left| \mathbf{x}\right| \leq r}x^{-1}\widehat{p}\left( 
\mathbf{x}\right) d^{3}\mathbf{x}\right] ^{2}\right| 0\right\rangle ,
\label{31c}
\end{eqnarray}

Here the operators $\widehat{\rho }$ and $\widehat{p}$ may depend on time,
but this dependence is not explicitly exhibited. It may be realized that the
quantum operators $\widehat{\rho }$ and $\widehat{p}$ appear always in
symmetrical ordering.

The next hypothesis refers to the correlations between the quantum
fluctuations at two different points but equal times (we do not need them at
different times).

\begin{proposition}
The correlations between the vacuum operators of density and pressure at two
space points and equal times depend only on the distance between the points,
that is they might be written 
\begin{eqnarray}
\left\langle 0\left| \widehat{\rho }_{vac}\left( \mathbf{x}\right) \widehat{%
\rho }_{vac}\left( \mathbf{y}\right) \right| 0\right\rangle  &=&f_{\rho \rho
}\left( \left| \mathbf{x-y}\right| \right) ,\;\left\langle 0\left| \widehat{p%
}_{vac}\left( \mathbf{x}\right) \widehat{p}_{vac}\left( \mathbf{y}\right)
\right| 0\right\rangle =f_{pp}\left( \left| \mathbf{x-y}\right| \right) , 
\nonumber \\
f_{\rho p}\left( \left| \mathbf{x-y}\right| \right)  &=&\frac{1}{2}%
\left\langle 0\left| \widehat{\rho }_{vac}\left( \mathbf{x}\right) \widehat{%
p_{vac}}\left( \mathbf{y}\right) +\widehat{p}_{vac}\left( \mathbf{y}\right) 
\widehat{\rho }_{vac}\left( \mathbf{x}\right) \right| 0\right\rangle .
\label{33}
\end{eqnarray}
\end{proposition}

This hypothesis is consistent with the homogeneity and isotropy at large
scale (cosmological principle). Actually the distance between points should
involve the metric, but I may assume that the approximation of Minkowski
metric is good enough in this case (in fact $\left| A\left( \mathbf{x}%
,t\right) -1\right| <<1$, $\left| B\left( \mathbf{x},t\right) -1\right| <<1$
for any $\mathbf{x}$ and $t$.)

If I insert eqs.$\left( \ref{32a}\right) $ into eq.$\left( \ref{31b}\right) $
I get, taking eqs.$\left( \ref{0c}\right) $ and $\left( \ref{33}\right) $
into account, 
\begin{eqnarray}
A\left( r,t\right)  &\simeq &1+\frac{8\pi }{3}G\rho _{mat}\left( t\right)
r^{2}+\frac{64\pi ^{2}}{9}G^{2}\rho _{mat}\left( t\right) ^{2}r^{4} 
\nonumber \\
&&+4G^{2}r^{-2}\int_{\left| \mathbf{x}\right| \leq r}d^{3}\mathbf{x}%
\int_{\left| \mathbf{y}\right| \leq r}d^{3}\mathbf{y\,\;}f_{\rho \rho
}\left( \left| \mathbf{x-y}\right| \right) ,  \label{32d}
\end{eqnarray}
As expected from eqs.$\left( \ref{0c}\right) $ the leading contribution from
the vacuum fluctuations is of order $G^{2}$. It is possible to perform the
angular integrals if I define the new function $L_{\rho \rho }(x,y)$ by 
\begin{equation}
L_{\rho \rho }(x,y)\equiv xy\int_{0}^{\pi }f_{\rho \rho }\left( \left| 
\mathbf{x-y}\right| \right) \sin \theta d\theta =\int_{\left| x-y\right|
}^{x+y}f_{\rho \rho }\left( z\right) zdz,  \label{32f}
\end{equation}
where 
\[
z\equiv \left| \mathbf{x-y}\right| =\sqrt{x^{2}+y^{2}-2xy\cos \theta },
\]
Thus I get for the last term of eq.$\left( \ref{32d}\right) $%
\begin{equation}
A_{vac}=8\pi ^{2}G^{2}I_{1},\;I_{1}\equiv 4r^{-2}\int_{\mathbf{0}%
}^{r}xdx\int_{\mathbf{0}}^{r}ydy\;L_{\rho \rho }(x,y).  \label{33c}
\end{equation}

Similarly I define 
\begin{equation}
L_{\rho \rho }(x,y)\equiv xy\int_{0}^{\pi }f_{\rho \rho }\left( \left| 
\mathbf{x-y}\right| \right) \sin \theta d\theta ,\;L_{\rho \rho }(x,y)\equiv
xy\int_{0}^{\pi }f_{\rho \rho }\left( \left| \mathbf{x-y}\right| \right)
\sin \theta d\theta .  \label{32g}
\end{equation}
Thus inserting eqs.$\left( \ref{32a}\right) $ into eq.$\left( \ref{31c}%
\right) $ I obtain, taking eqs.$\left( \ref{0c}\right) $ and $\left( \ref{33}%
\right) $ into account, 
\begin{equation}
B(r,t)\simeq 1+\frac{4\pi }{3}G\rho _{mat}\left( t\right) r^{2}+\frac{8\pi
^{2}}{3}G^{2}\rho _{mat}\left( t\right) ^{2}r^{4}+B_{vac},  \label{33d}
\end{equation}
where 
\begin{equation}
B_{vac}=8\pi ^{2}G^{2}\sum_{k=2}^{6}I_{k},  \label{33e}
\end{equation}
and the integrals $I_{k}$ are 
\begin{eqnarray}
I_{2} &=&4\int_{0}^{r}x^{-3}dx\int_{0}^{x}udu\int_{0}^{x}vdv\;L_{\rho \rho
}(u,v),  \nonumber \\
I_{3} &=&4\int_{0}^{r}x^{-1}dx\int_{0}^{x}udu\;L_{\rho p}(x,u),  \nonumber \\
I_{4}
&=&2\int_{0}^{r}x^{-2}dx\int_{0}^{r}y^{-2}dy\int_{0}^{x}udu\int_{0}^{y}vdv%
\;L_{\rho \rho }(u,v),  \nonumber \\
I_{5} &=&2\int_{0}^{r}x^{-2}dx\int_{0}^{x}udu\int_{0}^{r}dv\;L_{\rho p}(u,v),
\nonumber \\
I_{6} &=&2\int_{0}^{r}dx\int_{0}^{r}dy\;L_{pp}(x,y).  \label{34a}
\end{eqnarray}

We cannot proceed further until we fix the functions $L(x,y)$ or, what is
equivalent, the functions $f\left( \left| \mathbf{x-y}\right| \right) .$This
is made with our next hypothesis, which is justified as follows. By
comparison of the quantity $A_{vac}$, eq.$\left( \ref{33c}\right) ,$ with eq.%
$\left( \ref{metric}\right) $ we see that agreement requires that the
integral $I_{1}$ should be proportional to $r^{2}$. Similar scaling is
needed for the integrals $I_{2}$ to $I_{6}$. As a consequence the functions $%
f_{\rho \rho }\left( \left| \mathbf{x-y}\right| \right) ,f_{\rho p}\left(
\left| \mathbf{x-y}\right| \right) $ and $f_{pp}\left( \left| \mathbf{x-y}%
\right| \right) ,$ should scale with distance as $r^{-2}$. This leads to the
following assumption.

\begin{proposition}
The correlations between components of the vacuum energy-momentum tensor at
two different point and equal times should be proportional to the inverse of
the square of the distance between the points. That is 
\begin{eqnarray}
f_{\rho \rho }\left( \left| \mathbf{x-y}\right| \right)  &=&C_{\rho \rho
}\left| \mathbf{x-y}\right| ^{-2},\;f_{\rho p}\left( \left| \mathbf{x-y}%
\right| \right) =C_{\rho p}\left| \mathbf{x-y}\right| ^{-2},  \nonumber \\
f_{pp}\left( \left| \mathbf{x-y}\right| \right)  &=&C_{pp}\left| \mathbf{x-y}%
\right| ^{-2},  \label{35}
\end{eqnarray}
where $C_{\rho \rho },C_{\rho p}$ and $C_{p}$ are constant quantities, that
is independent of $\mathbf{x}$ and $t$.
\end{proposition}

After that the calculation of the integrals $I_{1}$ to $I_{6}$ is
straightforward although lengthy. I get 
\begin{eqnarray}
I_{1} &=&2C_{\rho \rho }r^{2},\;I_{2}=C_{\rho \rho }r^{2},\;I_{3}=2C_{\rho
p}r^{2},\;I_{4}=\left( \frac{4}{3}\log 2-\frac{1}{3}\right) C_{\rho \rho
}r^{2},\;  \nonumber \\
I_{5} &=&\left( \frac{5}{3}\log 2-\frac{4}{3}\right) C_{\rho
p}r^{2},\;I_{6}=\log 2C_{pp}r^{2}.  \label{36}
\end{eqnarray}
Agreement of $A_{vac}$, eq.$\left( \ref{33c}\right) ,$ and $B_{vac}$, eq,$%
\left( \ref{33e}\right) ,$ with eq.$\left( \ref{metric}\right) $ will be
obtained if 
\begin{equation}
16\pi ^{2}G^{2}C_{\rho \rho }=\frac{8\pi }{3}G\rho _{DE}\Rightarrow \rho
_{DE}=6\pi GC_{\rho \rho }.  \label{37}
\end{equation}
\begin{equation}
3\pi G\left[ \frac{2}{3}\left( 2\log 2+1\right) C_{\rho \rho }+\frac{1}{3}%
\left( 5\log 2+4\right) C_{\rho p}+2\log 2C_{pp}\right] =-\rho _{DE}.
\label{38}
\end{equation}
This is only possible if the correlations of the quantum vacuum
fluctuations, given by eqs,$\left( \ref{33}\right) $ and $\left( \ref{35}%
\right) ,$ fulfil 
\begin{equation}
\frac{4}{3}\left( \log 2+2\right) C_{\rho \rho }+\frac{1}{3}\left( 5\log
2+4\right) C_{\rho p}+2\log 2C_{pp}=0.  \label{39}
\end{equation}
In order to fix the quantities $C_{\rho \rho },C_{\rho p}$ and $C_{p}$ I
need an additional assumption, which I will make as follows.

\begin{proposition}
The density correlation and the pressure correlation are equal.
\end{proposition}

Thus I obtain a rather simple relation between the said quantities, namely 
\[
C_{\rho \rho }=C_{pp}=-\frac{1}{2}C_{\rho p}. 
\]

In summary the calculation \textit{suggests }that the ``dark'' energy (or
mass) density, $\rho _{DE}$, and pressure, $p_{DE}$, are fictitious but the
curvature of space-time is real and it is the same that would be produced by
a mass density and a pressure as in eq.$\left( \ref{1}\right) .$ The value
of the mass density, $\rho _{DE}$, may be obtained as a product of
Newton\'{}s constant, $G$, times some factor, $Q$, which depends on the
properties of the vacuum quantum fields, likely those of the standard model
of elementary particles. We might estimate the order of the parameter $Q$ by
means of a dimensionally correct combination of the Planck constant, $
\rlap{\protect\rule[1.1ex]{.325em}{.1ex}}h%
$, the speed of light, $c$, and a typical mass of elementary particles, $m$.
Consequently, in order that $\rho _{DE}$ has dimensions of energy density,
we shall assume 
\begin{equation}
\rho _{DE}\sim G\frac{m^{6}c^{2}}{
\rlap{\protect\rule[1.1ex]{.325em}{.1ex}}h%
^{4}}.  \label{40}
\end{equation}
Unfortunately eq.$\left( \ref{40}\right) $ is very sensitive to the actual
value of the unknown mass $m$. Likely the mass lies somewhere between the
electron and the proton mass, which gives 
\[
10^{-35}kg/m^{3}\lesssim \rho _{DE}\lesssim 10^{-24}kg/m^{3},
\]
a rather wide interval. In any case the results of our calculations are
compatible with the observed value, eq.$\left( \ref{1}\right) ,$ and far
from the value eq.$\left( \ref{2}\right) .$ Consequently our results are
consistent with the assumption that dark energy is just a fictitious energy
and pressure appropriate in order to parametrize the curvature of space-time
due to quantum vacuum fluctuations.

\end{document}